\documentclass[nofootinbib,prd,12pt,preprint,longbibliography,aps]{revtex4-1}

\usepackage{graphicx}
\usepackage{epstopdf}
\usepackage{amsmath}
\usepackage{amssymb}
\usepackage{amsfonts}
\usepackage{amsthm}
\usepackage{color}
\usepackage{array}
\usepackage{verbatim}
\usepackage{url}
\usepackage{graphicx}
\usepackage{braket}
\usepackage{epsfig}
\usepackage[
      colorlinks=true,
      linkcolor=blue,
      urlcolor=magenta,
      filecolor=blue,
      citecolor=red,
      pdfstartview=FitV,
      pdftitle={},
      pdfauthor={},
      pdfsubject={},
      pdfkeywords={},
      pdfpagemode={},
      bookmarksopen=true
	  ]{hyperref}

\newcommand{\abs}[1]{\left|#1\right|}

\def\frac#1#2{{#1\over #2}}

\def\be{\begin{equation}}
\def\ee{\end{equation}}
\def\ba{\begin{eqnarray}}
\def\ea{\end{eqnarray}}

\numberwithin{equation}{section}
\begin{document}

\title{Mass inflation and strong cosmic censorship in a nonextreme BTZ black hole\\}

\author{Srijit Bhattacharjee$^{a}$\footnote{srijuster@gmail.com}}
\affiliation{${^a}$Indian Institute of Information Technology (IIIT), Allahabad, \\
Deoghat, Jhalwa, Uttar Pradesh, India 211015}

\author{Shailesh Kumar$^{a}$\footnote{shaileshkumar.1770@gmail.com }}
\affiliation{${^a}$Indian Institute of Information Technology (IIIT), Allahabad, \\
	Deoghat, Jhalwa, Uttar Pradesh, India 211015}

\author{Subhodeep Sarkar$^{a}$\footnote{subhodeep.sarkar1@gmail.com }}
\affiliation{${^a}$Indian Institute of Information Technology (IIIT), Allahabad, \\
	Deoghat, Jhalwa, Uttar Pradesh, India 211015}

\date{\today}

\begin{abstract}
We study the phenomena of mass inflation using the Ori model for a rotating BTZ black hole that is sufficiently far from extremality, and show that the right Cauchy horizon ($\mathcal{CH}^+_R$) of the BTZ black hole becomes singular. Motivated by the recent analysis of Dias, Reall, Santos [J. High Energy Phys. 12 (2019) 097], we choose the retrograde quasinormal modes to govern the decay of perturbations exterior to the black hole. The resulting model captures the violation of the strong cosmic censorship conjecture near extremality. On the other hand, far from extremality, the $\mathcal{CH}^+_R$ of a BTZ black hole develops a weak null singularity. Our analysis shows a slowly rotating BTZ black hole will respect strong cosmic censorship.
\end{abstract}

\maketitle

\section{Introduction}\label{introduction}

The inner horizon of a rotating or a charged black hole is a place beyond which physics becomes unpredictable. In the maximal analytic extension of Kerr and Reissner-Nordstr\"{o}m black holes, the inner horizon serves as a boundary of future Cauchy development and is hence termed as the Cauchy horizon. The failure of determinism in any physical spacetime is a pathology. Therefore the existence of a region behind the Cauchy horizon as a part of regular spacetime poses a serious problem. However, Penrose and Simpson argued that the right Cauchy horizon ($\mathcal{CH}^+_R$) of a charged black hole is a surface of  \emph{infinite blue-shift} which will make $\mathcal{CH}^+_R$ unstable due to the amplified dynamic perturbations of the metric.  This motivated Penrose to propose the strong cosmic censorship conjecture (SCC) which says that for generic asymptotically flat initial data, the maximal Cauchy development of a ``reasonable" Einstein-matter system is future inextendable as a suitably regular Lorentzian manifold \cite{Penrose:1968ar, Simpson:1973ua}. Therefore if one allows perturbations to the initial data, then the evolution of a generic Einstein-matter system will render $\mathcal{CH}^+_R$\footnote{Our study will be confined only to the instability of the right portion of Cauchy horizon ($\mathcal{CH}^+_R$) in Penrose diagram. See Figure \ref{ori-model_FIGURE}. } unstable, and the spacetime will be sealed beyond it. It is now a well-established fact that $\mathcal{CH}^+_R$ of asymptotically flat ($AF$) black holes are singular \cite{McNamara499,chandra,Poisson:1989zz,Poisson:1990eh,Dafermos:2003wr}. Although these singularities at $\mathcal{CH}^+_R$ are not strong enough to stop one from extending the spacetime as a continuous metric beyond $\mathcal{CH}^+_R$, but the derivative of the metric, in general, cannot be continuously extendable across $\mathcal{CH}^+_R$ \cite{Poisson:1989zz,Poisson:1990eh, Ori:1991zz,Dafermos:2003wr,Dafermos:2017dbw}. More sophisticated analyses have prompted the formulation of a more precise version of SCC, which says that one cannot extend the spacetime beyond $\mathcal{CH}^+_R$ as a continuous metric with locally square-integrable Christoffel symbols  \cite{Christodoulou:2008nj,Christodoulou2009,Dafermos:2012np}. This version of SCC does not rule out a macroscopic observers' passage, without being torn apart, while approaching the $\mathcal{CH}^+_R$, but provides a well-defined notion of inextendability. So generically, when perturbations are present, $\mathcal{CH}^+_R$ is replaced by a weak null singular surface. This picture has remained true for all $AF$ black holes in Einstein gravity. 

%%%
%%
The Poisson-Israel \emph{mass inflation} model \cite{Poisson:1989zz,Poisson:1990eh} provided one of the earliest inroads toward finding the singularity at $\mathcal{CH}^+_R$ of a charged black hole in $AF$ spacetime. Mass inflation is characterized by an unbounded increase of the quasilocal Hawking mass of a perturbed black hole, triggering a singularity at $\mathcal{CH}^+_R$. Shortly after this discovery, Ori \cite{Ori:1991zz} was able to considerably simplify the model while retaining its capacity to paint a vivid physical picture \cite{Ori:1991zz}. The Ori model could also capture the fact that the nature of singularity at $\mathcal{CH}^+_R$ is a \emph{weak null singularity}, i.e., the tidal forces on a macroscopic body are infinite, but the tidal distortions remain finite as it approaches $\mathcal{CH}^+_R$. This model was further confirmed by numerical simulations \cite{Brady:1995ni, Burko:1997zy, Hod:1998gy} and has gained considerable support from rigorous mathematical analyses \cite{Dafermos:2003wr,Dafermos:2003yw,Dafermos:2004jp}. 

%%%
%%%

The extension of SCC to black hole spacetimes possessing a positive cosmological constant (the de Sitter space) has been well explored, anticipating a violation of the SCC due to the presence of a cosmological horizon.  In fact, it was proposed that presence of a redshift effect of the infalling perturbations inside a  Reissner-Nordstr\"{o}m-de Sitter (RN-$dS$) black hole, associated with the cosmological horizon, makes the backreaction weaker in the near-extreme RN-$dS$ black hole (see \cite{Chambers:1997ef} for a review of earlier works including the mass inflation model). Later, it was shown that the backreaction of classical perturbations of a scalar field is strong enough to render $\mathcal{CH}^+_R$ of RN-$dS$ black hole unstable  \cite{Brady:1998au}. Recently, more detailed analyses have revealed there is indeed a violation of SCC in RN-$dS$ spacetime near the extremality  \cite{Cardoso:2017soq,Dias:2018etb,Luna:2018jfk} although the same is not true for Kerr-$dS$ black hole \cite{Dias:2018ynt}. In a recent note \cite{Hollands:2019whz}, it has been shown that although classically, there seems to be a violation of SCC in RN-$dS$ black holes, but quantum effects will restore the singular behavior at $\mathcal{CH}^+_R$ rescuing the SCC. 

%%%%
%%%
At present, investigations into the violation of SCC for black holes in anti-de Sitter ($AdS$) space (solution of Einstein equations with a negative cosmological constant) have gained much attention. In a recent detailed study by Dias, Reall and Santos \cite{Dias:2019ery}, it has been unveiled that a near-extremal rotating BTZ black hole violates SCC\footnote{BTZ black holes are $(2+1)$ dimensional solutions of Einstein equation with negative cosmological constant \cite{ Banados:1992wn,Banados:1992gq, Carlip:1995qv}.}, correcting some previous results  \cite{Chan:1994rs, Husain:1994xa, Kraus:2002iv,Levi:2003cx,Balasubramanian:2004zu}. In particular, this result is against the expectation of the classical analysis presented in \cite{Balasubramanian:2004zu}. This violation is an artifact of a remarkable coincidence that certain \emph {exterior prograde} quasinormal match modes with certain \emph{interior} quasinormal modes (QNMs) of the black hole \cite{Dias:2019ery}. This indicates the nonsmooth behavior of the stress-energy tensor of a test scalar field is not governed by the slow-decaying prograde mode but a faster-decaying \emph{retrograde} mode near  $\mathcal{CH}^+_R$. In another study, this result has been supported by a detailed numerical analysis in the initial value formulation setup \cite{Pandya:2020ejc}. 

%%%%%%%%%%%%%%%%%%%%%
In \cite{ Dias:2019ery}, it was also shown that quantum effects near $\mathcal{CH}^+_R$ do not enforce SCC for a near-extreme BTZ black hole. The violation of SCC in a BTZ black hole (in the semiclassical regime) has recently been supported by Hollands \textit{et al}. \cite{Hollands:2019whz}. From the $AdS/CFT$ perspective, recent works \cite{Papadodimas:2019msp, Balasubramanian:2019qwk} have also supported the findings of \cite{ Dias:2019ery}. However, it has been argued in \cite{Emparan:2020rnp}, the quantum stability of $\mathcal{CH}^+_R$ in BTZ would be lost once we consider beyond linear order backreaction of the probe field. It is not clear from these recent works if quantum effects near $\mathcal{CH}^+_R$ will always be small irrespective of the fact whether the black hole is close to extremality. But there exists a sufficiently wide parameter space where the classical instability of $\mathcal{CH}^+_R$ will prevail, and SCC is respected \cite{Pandya:2020ejc}. 

%%%%
%%%
In this paper, we study the mass inflation scenario for a BTZ black hole by implementing the Ori model \cite{Ori:1991zz}. The dynamics of a probe field inside a rotating or a charged black hole crucially depends on the rate of decay of the perturbing field at the exterior of a black hole. In $AF$ spacetimes, the perturbations of test fields decay with a power law tail at late times according to Price's law  \cite{Price:1971fb}. This slower decay triggers an instability at $\mathcal{CH}^+_R$ of the black hole. In BTZ spacetime, the late time decay of perturbations is governed by low-lying quasinormal modes (QNMs), so the decay is exponential. We have already mentioned that there are two types of QNMs in a BTZ black hole, the prograde (corotating with black hole) and retrograde (counterrotating) modes \cite{Birmingham:2001hc, Balasubramanian:2004zu, Dias:2019ery}. As shown in \cite{Dias:2019ery}, among these two, it is the faster decaying retrograde mode that plays a crucial role in determining the diverging behavior of  $\mathcal{CH}^+_R$. 

%%%%%%%%%%%%%%%%%%%%%%%%%%%%%%%%%%

In the mass inflation analysis, one would consider the lowest-lying QNMs to be the appropriate modes that govern the decay of infalling radiation at late times. In the case of BTZ black holes, it is the prograde mode that is the lowest or long-lived. However, to match the intricate dynamics and nonsmooth behavior of the test scalar field as depicted in \cite{Dias:2019ery}, we find surprisingly that it is the faster decaying retrograde modes give the correct decay rate for which the stress-energy tensor of a massless scalar field diverges exactly at the same rate obtained in \cite{Dias:2019ery}. Hence, although the slowest decaying prograde modes would have naively produced a mass inflation singularity for BTZ black hole, but it would have failed to capture the intricate classical dynamics discerned in \cite{Dias:2019ery}. The mass inflation model for BTZ black hole with the retrograde QNMs also produces the smooth behavior of the gradient of the probe field near $\mathcal{CH}^+_R$ and indicates the violation of strong cosmic censorship near extremality as found in \cite{Dias:2019ery}. Our characterization of singularity and smoothness of $\mathcal{CH}^+_R$ also exactly matches with that of \cite{Dias:2019ery}.  Near $\mathcal{CH}^+_R$, the diverging component of the stress-energy tensor in the mass inflation model behaves as
\begin{equation}
V^{-2(1-\beta)},
\end{equation}
where $V$ is a null Kruskal-like coordinate and $\beta$ is proportional to the longest-lived retrograde QNM for BTZ black hole. Clearly, for $\beta>1$, one finds that the stress-energy tensor becomes finite at $\mathcal{CH}^+_R$ indicating a violation of SCC. In a similar vein, we also show that the spin parameter $a$ of a BTZ black hole has to be less than $0.6$ for mass inflation to occur and respect SCC, a result that has recently been obtained via numerical simulation in \cite{Pandya:2020ejc}. In fact, a stronger bound, $a<0.38$, is obtained by demanding that only the Christodoulou version of SCC should be respected.  Remarkably, the heuristic mass inflation model can anticipate the intricate dynamics of the interior of a BTZ black hole.

%%%%%%

The organization of the paper is, therefore, as follows: In Sec. \ref{massinflation}, we depict a detailed study of the mass inflation model for a rotating BTZ black hole. We discuss the importance of choosing the retrograde QNMs in the context of classical backreaction induced at the $\mathcal{CH}^+_R$ and show the divergent behavior of Hawking mass at $\mathcal{CH}^+_R$. In Sec. \ref{regular_soln_section}, we obtain a regular solution near the mass inflation region. This has not been done earlier, and thus fills a long-standing gap in the literature. Next, with the regular metric near $\mathcal{CH}^+_R$, we construct a freely falling triad and show that the tidal distortions remain finite as a test body tries to cross $\mathcal{CH}^+_R$. We also show how the model captures the violation of SCC in near-extreme regime as obtained in \cite{Dias:2019ery} and \cite{Pandya:2020ejc}. In Sec. \ref{bound}, we determine an upper bound on the spin parameter of a BTZ black hole above which SCC is violated. In Sec. \ref{backreaction}, we include a short discussion on the quantum effects at $\mathcal{CH}^+_R$ of a BTZ black hole, and we finally conclude in Sec. \ref{discussion}, with a summary of our results, pointing out some future prospects.

\section{Mass Inflation in a Rotating BTZ Black Hole}\label{massinflation}

The mass inflation scenario is a simple and elegant model where one studies the backreaction of blue-shifted radiation near $\mathcal{CH}^+_R$. In this model, one mimics Penrose's blue-shifted radiation as an influx of null dust. This influx then backreacts with the black hole's spacetime and gives rise to an outflux inside the outer horizon of the hole. Due to the presence of the influx and outflux, the background geometry (which in our case is a BTZ black hole) changes to a set of BTZ-Vaidya solutions \cite{Poisson:1990eh}. One then computes the Hawking mass\footnote{See \cite{Szabados:2009eka} for a detailed exposition on Hawking mass and other local quantities in general relativity.} near $\mathcal{CH}^+_R$ in the presence of this cross-flow. It should be noted that an outflux is an essential ingredient in triggering the blow-up of the mass function at $\mathcal{CH}^+_R$; without it, one does not get an appreciable effect on the inner horizon. However, the detailed structure of outflux is not essential, its role is to separate the inner apparent horizon from $\mathcal{CH}^+_R$. This observation led Ori to model the outflux as an extremely short pulse of null radiation \cite{Ori:1991zz} and significantly reduced the mathematical complexity of the Poisson-Israel model. In the Ori model, we thus need to match two BTZ-Vaidya solutions along a delta function shell, and determine the behavior of the mass function of BTZ-Vaidya metric near $\mathcal{CH}^+_R$. The decay of the energy density (the ``mass function") of the ingoing radiation is an input in this model. In fact, the late time behavior of the influx outside and along the outer event horizon plays a crucial role in obtaining a singularity at $\mathcal{CH}^+_R$.

 The Ori model of mass inflation in a BTZ black hole was first studied by \cite{Chan:1994rs}. But their analysis \cite{Chan:1994rs} was lacking in one crucial aspect: they had assumed that the mass function decays at late times according to a power law similar to Price's law \cite{Price:1971fb}. However, the authors addressed this issue in \cite{Chan:1996yk}, noting that a power law tail is absent in BTZ black hole, and the true decay behavior will be exponential. Later, exact expressions of quasinormal modes (QNMs) of BTZ black holes were obtained in \cite{Birmingham:2001hc}. We intend to first analyse the mass inflation model for a BTZ black hole with the help of the exact expressions of QNMs. The initial part of this section shall have a substantial overlap with earlier analyses as we have reviewed certain things to coherently and meticulously present the results obtained in this study. 

In a $(2+1)$ dimensional spacetime with a negative cosmological constant $\Lambda=-1/L^2$ (where $L$ is the $AdS$ radius), the solution to the vacuum Einstein field equation is a BTZ black hole \cite{Banados:1992wn,Banados:1992gq, Carlip:1995qv} which is described by the metric
\be
ds^2=-N^2dt^2+N^{-2}dr^2+r^2\left(d\phi -(J/2r^2)dt\right)^2,
\ee
where $N^2(r)$ is the lapse function, given by
\begin{equation} \label{lapsefunc}
N^2(r)=\alpha(r) + \frac{J^2}{4r^2},
\quad\mathrm{and}\quad 
\alpha(r) = - M-\Lambda r^2. 
\end{equation}
In advanced Eddington-Finkelstein (EF) or  $\{v, r, \theta\}$ coordinates, this metric reads
\begin{equation}\label{btzm-basic}
ds^2= - \alpha(r) {\mathop{dv}}^2 + 2 \mathop{dv}\mathop{dr}- J\mathop{dv}\mathop{d \theta} + r^2 {\mathop{d \theta}}^2.
\end{equation}
The constants $M=(r_+^2 +r_-^2)/r^2L^2$ and $J=2r_+r_-/L$ are identified, respectively, as the mass and the angular momentum of the black hole, and the BTZ black hole has two horizons, an event horizon at $r=r_{+}$, and an inner horizon at $r=r_{-}$ where $r_{\pm}$ are the two solutions of $N^2(r)=0$.
The radius of outer and inner horizons are expressed as
\be \label{adef}
r_{\pm}=\sqrt{\dfrac{ML^2}{2}\left(1\pm \sqrt{1-a^2}\right)},~\mathrm{where}~ 
a=\dfrac{J}{ML}.
\ee

 The stress-energy tensor of the ingoing flux has the form
\begin{equation}
T_{\mu \nu}=\dfrac{\rho(v)}{r} \delta_{\mu v}\delta_{\nu v},
\end{equation}
and with this form of the stress-energy tensor, the Einstein equation gives the rotating \textit{BTZ-Vaidya} metric
\begin{equation} \label{btz_vaidya}
ds^2= - \alpha(v,r) {\mathop{dv}}^2 + 2 \mathop{dv}\mathop{dr}- J\mathop{dv}\mathop{d \theta} + r^2 {\mathop{d \theta}}^2,
\end{equation}
where $\alpha(v,r)= -m(v)-\Lambda r^2 =  N^2(v,r)-J^2/(4r^2)$. The mass function $m(v)$ should also satisfy the equation 
\begin{equation}\label{dmdv}
\dfrac{\mathop{dm}}{\mathop{dv}}=16 \pi G \rho(v),
\end{equation}
such that, if $\rho(v)$ decays at least as fast as $1/v$ for very large $v$, then $ \lim_{v \to \infty} m(v) = M$ where $M$ is a constant that can be identified as the mass of the black hole.

\begin{figure}
	\centering
	\includegraphics[width=75mm]{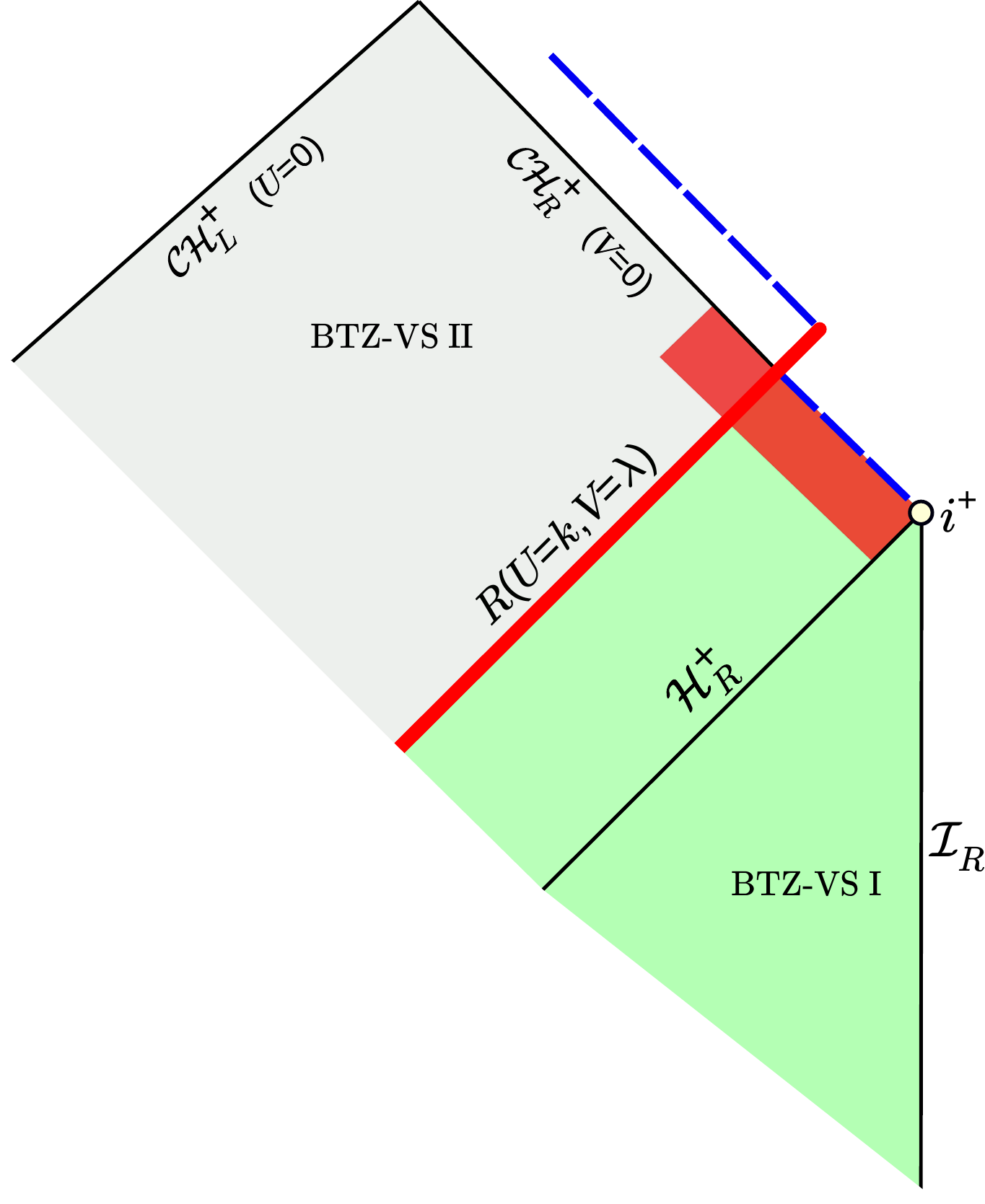}
	\caption{A schematic diagram of the Ori model for a rotating BTZ black hole: The thick red line corresponds to the null ring $R$ mimicking the outgoing radiation and the red rectangular strip is denoting the influx.  The dashed blue lines denote the inner apparent horizon. The future event horizon ($\mathcal{H}^+_R$), the right and left future Cauchy horizons ($\mathcal{CH}^+_R$ and $\mathcal{CH}^+_L$, respectively) have been also indicated in the diagram. $U, V$ are regular double null coordinates to be introduced later.}
	\label{ori-model_FIGURE}
\end{figure}

Now, this ingoing flux of massless particles will get backscattered after crossing the event horizon, and result in an outflux of massless particles. This outflux can be modeled as a thin outgoing null ring $\mathcal{R}$ \cite{Ori:1991zz, Chan:1994rs} in the usual $r-\phi$ coordinates to study the phenomena of mass inflation. We consider this ring $\mathcal{R}$ to be situated in a region between $\mathcal{H}^+_R$ and $\mathcal{CH}^+_L$ (see Fig. \ref{ori-model_FIGURE}), and the quantities outside the ring are labeled with a subscript $1$, and those inside the ring with a subscript $2$.

The Ori model, therefore, consists of two  BTZ Vaidya solutions (see Fig. \ref{ori-model_FIGURE}) in regions I (outside the ring) and II (inside the ring) that are to be matched across the null ring $\mathcal{R}$. The preliminary junction condition demands the components of the metric tensor to be continuous across $\mathcal{R}$, thereby, forcing the coordinate $r$ to be continuous. Moreover, since the ring has a vanishing surface tension \cite{Barrabes:1991ng, Ori:1991zz}, the same affine parameter $\lambda$ can be used on either side of the ring.

We can now proceed to write down the matching conditions. Let us first specify a function $R(\lambda)$ such that $2\pi R(\lambda)$ will give us the perimeter of the ring $\mathcal{R}$. Here, $\lambda$ is an affine parameter which is set to zero at  ($\mathcal{CH}^+_R$), and is negative below it, so $\lambda$ increases with time. From (\ref{btz_vaidya}), the nullity condition of the ring gives us
\begin{align} \label{nullity_condition}
- &\alpha(v,R(\lambda)) {\dot{v}}^2(\lambda) + 2  \dot{v}(\lambda)\dot{R}(\lambda)- J\dot{v}(\lambda)\dot{\theta}(\lambda,\chi) \nonumber \\ & + R^2(\lambda) \dot{ \theta}^2(\lambda,\chi)=0,
\end{align}
with $\chi$ being the intrinsic coordinate of the ring, and the overdot denoting derivatives with respect to $\lambda$.
The geodesic equation corresponding to the $\theta$ coordinate is 
\begin{equation} \label{theta_geodesic}
\dfrac{d}{d \lambda}\left(-J \dot{v} + 2 \dot{R}^2 \dot{\theta}^2\right)=0.
\end{equation}
This equation can be readily integrated to give
\begin{equation}  \label{theta_dot_eq}
\dot{\theta}=\dfrac{J}{2 R^2} \dot{v},
\end{equation}
where we have chosen $\theta(\lambda_0, \chi)= \chi_0$, a constant, for some initial value $\lambda_0$ of the affine parameter to set the constant of integration to zero.

In the above equations, we had suppressed the subscript $i=1,2$ on $v$. Restoring the subscript $i$ and using (\ref{theta_dot_eq}), the nullity condition (\ref{nullity_condition}) can be recast as
\begin{equation}\label{nullity2}
\dfrac{\dot{R}}{\dot{v_i}}= \dfrac{1}{2}\left(\alpha + \dfrac{J^2}{4R^2}\right)=\dfrac{1}{2}N^2(R(\lambda, v_i)),
\end{equation}

Now, the radial geodesic equation
using (\ref{theta_dot_eq}), can be written as
\begin{flalign}
\ddot{v}_i &= \dfrac{1}{2}\dot{v}_i^2\left( \dfrac{\partial}{\partial r} \left( \Lambda r^2 + m(v_i)\right) - \dfrac{\partial}{\partial r} \left(\dfrac{J^2}{4r^2}\right)\right)\Bigr|_{r=R(\lambda)} \nonumber\\ &=-\dfrac{1}{2}\dot{v}_i^2 (\partial_r N^2)|_{\substack{r=R(\lambda)\\v=v_i(\lambda)}} \label{r_geodesic_eq}.
\end{flalign}
Defining 
\begin{equation} \label{z_eq}
z_i(\lambda)= \dfrac{2R(\lambda)}{\dot{v}_i (\lambda)},
\end{equation}
the nullity condition (\ref{nullity2}) can be put in the following form:
\begin{equation}\label{nullity3}
\dfrac{z_i(\lambda)}{2R(\lambda)}\dot{R}(\lambda) = \mathcal{M(R(\lambda))}- m_i(v_i(\lambda)),
\end{equation}
where $\mathcal{M}(r)$ denotes,
\begin{flalign}
\mathcal{M}(r) & =m_i(v_i(\lambda))+ \dfrac{1}{2}N^2(r, v_i(\lambda)) \label{curly_M_def1} = \Lambda r^2 + \dfrac{J^2}{4r^2}. 
\end{flalign}

We further rewrite (\ref{z_eq}) as $z_i/2R=1/\dot{v_i}$, and differentiate it with respect to $\lambda$ to cast the radial geodesic equation (\ref{r_geodesic_eq}) as
\begin{align} \label{z_eq_derivative}
\dfrac{\mathop{d}}{\mathop{d\lambda}} \left[\dfrac{z_i}{2R}\right] = \dfrac{1}{2}\partial_r N^2(v,r)\Big|_{\substack{r=R(\lambda)\\v=v_i(\lambda)}} = \partial_r \mathcal{M}(r)|_{{r=R(\lambda)}},
\end{align}
where we have taken partial derivative of (\ref{curly_M_def1}) with respect to $r$ to write down the last equality.

Finally, using (\ref{z_eq}), (\ref{nullity3}), and (\ref{z_eq_derivative}), we can write down the three matching equations
\begin{flalign}
 &v_i(\lambda) = \int^{\lambda}\dfrac{2R(\lambda ')}{z_i(\lambda ')} \mathop{d \lambda '} \label{matching_v}, \\
&m_i(v_i(\lambda))= \mathcal{M}(R(\lambda)) - \dfrac{z_i \dot{R}(\lambda)}{2R(\lambda)} \label{matching_m}, \\
&z_i(\lambda)=2R(\lambda)\left[Z_i + \int^{\lambda} \partial_r \mathcal{M}(R'(\lambda ')) \mathop{d \lambda '}\right], \label{matching Z}
\end{flalign}
where we have obtained (\ref{matching Z}) by integrating (\ref{z_eq_derivative}) with respect to $\lambda$; we have also ignored the constant of integration in (\ref{matching_v}) since adding a constant to the $v_i$ coordinate amounts to just shifting the origin of the coordinate system. The $Z_i$'s are constants of integration and the subscript $i=1,2$ denotes the respective region in which the quantity is defined. We note that after specifying the function $R(\lambda)$, along with the constants $Z_i$'s, the matching equations (\ref{matching_v}-\ref{matching Z}) will completely determine the rotating BTZ-Vaidya solution on either side of the null ring.

\subsection{Late time behavior of the mass function}\label{late_time_behavior}

The $\lambda$ dependent mass of the ring can be obtained from (\ref{matching_m}) and (\ref{matching Z}), and is given by
\begin{equation} \label{mass_eq}
\Delta m(\lambda)= m_2(\lambda) - m_1 (\lambda) = \left( Z_1-Z_2 \right) \dot{R}(\lambda).
\end{equation}

To determine the constants $Z_i$'s, we note that: 
 $\mathcal{CH}^+_R$ corresponds to the limit $v_1 \to \infty$ in region I, with $\lambda=0$ on $\mathcal{CH}^+_R$. This implies that the derivative of $v_1$ (with respect to $\lambda$) will also begin to blow up near $\mathcal{CH}^+_R$. So, using (\ref{matching_v}) and arguments similar to those in  \cite{Ori:1991zz,Bhattacharjee:2016zof,Chan:1994rs}, we find $Z_1=0$ and $Z_2>0$\footnote{Since we are considering an outgoing ring $\mathcal{R}$, the value of $\dot{R}(\lambda)$ is negative inside the black hole, so (\ref{mass_eq2}) would imply ${Z_2 > 0}$ for the energy of the ring to be positive.}.

Therefore, using (\ref{matching_v}) near $\mathcal{CH}^+_R$, we get
\begin{equation}\label{asym_eq}
|\lambda| \approx  e^{- \kappa_{-} v_1}
\quad\mathrm{and}\quad 
v_2 \approx \dfrac{2}{Z_2} \lambda
,\end{equation}
where 
\begin{equation} \label{kappa}
\kappa_{-} := -\dfrac{ \mathop{d \mathcal{M}(r)}}{\mathop{dr}}\Bigr|_{r=r_-} = - \dfrac{1}{2}\partial_r N^2(r, v_1 \to \infty) |_{r=r_-}.
\end{equation}

We see that $v_2$ is linearly proportional to the affine parameter $\lambda$ near the $\mathcal{CH}^+_R$. So, in region II, $\mathcal{CH}^+_R$  is located at $v_2=0$, since $\lambda=0$ on $\mathcal{CH}^+_R$ . Finally,Eq. (\ref{mass_eq}) can now be written as
\begin{equation} \label{mass_eq2}
\Delta m (v_i(\lambda)) = -Z_2 \dot{R}(\lambda).
\end{equation}

To obtain a solution of above equation, we consider the geometry near $\mathcal{CH}^+_R$  to be a slightly perturbed version of the rotating BTZ-Vaidya solution. Using the nullity equation (\ref{nullity2}), we can write

\begin{equation} \label{radial_eq}
\dfrac{\mathop{dR}(v_1)}{\mathop{dv_1}}=\mathcal{M}(R(v_1)) - m_1(v_1).
\end{equation}
Next, we expand $\mathcal{M}(R(v_1))$ around $r_{-}$ as
$ R(v_1)=r_{-} + \delta R(v_1)$, and obtain

\begin{equation} \label{radial_eq2}
\dfrac{\mathop{d}}{\mathop{d v_1}} \delta R(v_1) \approx \left( -\kappa_{-} \delta R + \delta m_1(v_1)\right).
\end{equation}
In writing down the above expression, we have used the fact that the final mass $m_f$ of the black hole as measured by an observer in region I, after it has absorbed all the infalling radiation, is given by
\begin{equation} \label{mass_func2}
m_f=\mathcal{M}(r_{-})=m_1(v_1)+ \delta m_1(v_1).
\end{equation}
Here, the term $\delta m$ is understood to be the ``mass" associated with the radiative tail that dominates the late time behavior of the infalling radiation flux. 

\subsubsection{Quasinormal modes of a BTZ black hole and stability-instability of $\mathcal{CH}^+_R$}\label{QMN}

The late time solution of the radial equation (\ref{radial_eq2}) clearly depends on the behavior of the mass function (\ref{mass_func2}). So we must specify the late time behavior $\delta m$ of the influx before proceeding further. As mentioned earlier, in general, a power law tail is not present in $AdS$ space \cite{Horowitz:1999jd}\footnote{If a BTZ black hole is formed in a decoupling limit of a higher-dimensional asymptotically flat black hole, a power law tail may arise.}. In this case, it is more appropriate to  consider quasinormal modes (QNMs) of $AdS$ black holes, which have been extensively studied in \cite{Horowitz:1999jd,Birmingham:2001hc, Festuccia:2008zx, Berti:2009kk,Konoplya:2011qq, Dias:2019ery}. In $d=4$, due to the \emph{stable trapping} phenomena of null geodesics for Kerr-$AdS$ black hole, there exists a much slower (logarithmic) decay of perturbations at late time \cite{Holzegel:2011uu}. This essentially triggers a more rapid inflation of mass parameter than $AF$ space \cite{Bhattacharjee:2016zof,Kehle:2018zws}, and leads to the development of a weak null singularity at $\mathcal{CH}^+_R$. However, the stable trapping of null geodesics is absent in a BTZ black hole. Therefore the decay of perturbations should be dictated by long lived QNMs.  

 For a massless scalar field perturbation, the exact expressions of QNMs are given by \cite{Birmingham:2001hc, Dias:2019ery}

\begin{align}\label{quasinormal} 
&\omega_p= \frac{ q}{L} - 2i\dfrac{r_{+} - r_{-}}{L^2}(n+1) ,
\quad\mathrm{and}\quad \nonumber\\
&\omega_r  = -\frac{ q}{L} - 2i\dfrac{r_{+} + r_{-}}{L^2}(n+1),
\end{align}
where $q$ is azimuthal number and $n$ is the overtone number. 

It is easy to see that the imaginary part of prograde modes are lower than retrograde modes. Therefore, a naive guess would be to consider the lowest-lying $\omega_p$ modes as input for the late time decay of mass function or, $\delta m(v)$ as $v\to \infty$. However, as shown in \cite{Dias:2019ery}, it is the faster retrograde mode that dictates any nonsmooth behavior of a scalar perturbation near $\mathcal{CH}^+_R$. We will come to this fact later, for the time being, let us define two different lowest lying QNMs $\omega_{p,r}$ by setting $q=0$ and $n=0$ as
\begin{equation} \label{quasinormal2}
\omega^{I}_{p,r}= -2\dfrac{r_{+}\mp r_{-}}{L^2}=-\zeta_{p,r} \kappa_{+},
\end{equation}
where $\zeta_{p,r}$ are dimensionless constants and $\kappa_{+}$ is the surface gravity of the event horizon of the final stationary black hole. We display the expressions of $\zeta_{p,r}$ and surface gravities ($\kappa_{\pm}$) below

\begin{equation}
\zeta_{p,r}=\frac{2r_{+}}{r_+ \pm r_-}, 
\quad\mathrm{and}\quad  \kappa_{\pm}=\frac{r_{+}^2-r_{-}^2}{L^2r_{\pm}}. \label{alpha-kappa}
\end{equation}
The asymptotic form of $\delta m(v_1)$, should be such that it is consistent with Eq. (\ref{dmdv}). The amplitude of the infalling radiation is modelled as perturbing test scalar field. Hence if the scalar field decays like $e^{-\zeta_{p,r} \kappa_{+} v_1}$, then the energy density of the massless field or flux, $\propto \dfrac{dm}{dv_1}$, should go as square of the amplitude, 
$e^{-2\zeta_{p,r} \kappa_{+} v_1}$ \cite{Poisson:1990eh}. This leads to
\begin{equation} \label{asymp_m1}
\delta m_{p,r}(v_1) \propto e^{2\omega^I_{p,r} v_1} = e^{ -2\zeta_{p,r} \kappa_{+} v_1}.
\end{equation}

\subsubsection{The divergent mass function}\label{diverging_mass_section}
If we solve the radial equation (\ref{radial_eq2}) with (\ref{asymp_m1}) we get
\begin{equation}
\delta R_{p,r}(v_1)= C_1 e^{-\kappa_{-}v_1}+ \dfrac{C_2}{\kappa_{-} - 2\zeta_{p,r} \kappa_{+}}e^{- 2\zeta_{p,r} \kappa_{+} v_1},
\end{equation}
where $C_1$ and $C_2$ are the constants of integration. The mass function in region II can now be obtained using (\ref{mass_eq2}) as
\begin{equation}\label{mass_diverging2}
m_2^{p,r}(v_2) \approx \Delta m_{p,r}(v_2) \propto |v_2|^{-\left(1-2\dfrac{ \zeta_{p,r} \kappa_{+}}{\kappa_{-}}\right)}.
\end{equation}

Note that for any nonextreme black hole: $\kappa_- >\kappa_+$. Now, we observe that if the black hole is sufficiently far from extremality, i.e., if $r_+-r_->>0$, both the  expressions $m_2^{p,r}(v_2)$ diverge as $\mathcal{CH}^+_R$ is approached i.e. as $v_2\to 0$. This is apparent from the fact $\zeta_{p,r}<1$ if one stays away from extremality.  However, as already discussed, we now show, the  mass function in terms of retrograde mode exactly produces the desired behavior. Recall the definition of $\beta$ for a massless field
 
\be\beta=\dfrac{2r_-}{r_+- r_-}.\ee 

A careful glance at the fraction $\dfrac{\zeta_r\kappa_+}{\kappa_-}$ reveals
\be \beta=\zeta_r\frac{\kappa_+}{\kappa_-}. \ee

%Therefore 
%\be m_2^r(v_2) \propto |v_2|^{-(1-\beta)}. \label{mretro}\ee

 Therefore
\be m_2^r(v_2) \propto |v_2|^{-(1-2\beta)}. \label{mretro}\ee

 Hence the mass function diverges if $\beta <1/2$, which means if one is far from extremality the right Cauchy horizon ($\mathcal{CH}^+_R$) becomes unstable. On the other hand, if $\beta>1/2$, or there is very fast decay of $\delta m$, the classical backreaction becomes zero at the $\mathcal{CH}^+_R$. So, the SCC is violated.\footnote{In terms of prograde mode, $\beta_p=\frac{2r_-}{r_++r_-}=1$ only in the exact extreme limit. Therefore, a violation of SCC will occur  not before the black hole is reached to its extremality. This is not the picture that we get from the recent studies \cite{Dias:2019ery,Pandya:2020ejc}.} This is the same conclusion obtained in \cite{Dias:2019ery} from the divergence properties of stress-energy tensor of the probe scalar field. It should also to be noted, this condition on $\beta$ indicates that the Christodoulou version of SCC is respected for retrograde modes. 
 
 Let us now compare these results with actual behavior of the stress-energy tensor of a test scalar field $\phi$ computed in \cite{Dias:2019ery}. The nonsmooth part of scalar field near the $\mathcal{CH}^+_R$ is expressed in terms of EF type coordinates. The scalar field has the following fall-off near 
 $\mathcal{CH}^+_R$ (see Eq. 3.46 of \cite{Dias:2019ery})
 \begin{equation}
     \phi \approx z^{\beta},
 \end{equation}
 where $z$ is defined as 
 \[z=\dfrac{r^2-r_{-}^2}{r^2_+-r_-^2}.\]
 
 Since the stress-energy tensor of a scalar field contains square of its derivative, we can easily see near $\mathcal{CH}^+_R$ the divergent behavior of stress-energy tensor is given by
 \begin{equation}
   T_{zz} \approx z^{2(\beta-1)}.
 \end{equation}
 As $z\to 0$ at $\mathcal{CH}^+_R$, we can see a divergence will set in if $\beta <1$. We can convert $z$ to null coordinate to directly compare with the behavior of local mass function. Near $\mathcal{CH}^+_R$, $z$ behaves as $z\sim e^{-2\kappa_- r_*}$ \cite{Balasubramanian:2004zu}, where $r_*$ is the radial tortoise coordinate. Therefore in terms of regular Kruskal like coordinates $U=-e^{\kappa_- u},\,V=-e^{-\kappa_- v}$ \cite{Dias:2019ery}, the behavior of stress-energy tensor takes the form

 \be T_{VV}\approx  |V|^{2(\beta-1)} \label{stfalloff}.\ee
 
 Since $v_2\sim \lambda$, it is Kruskal like in the region II (see next section), so $v_2\propto V$. Now
 a direct calculation of the Einstein tensor with the metric (\ref{btz_vaidya}) in region II yields the following leading divergent behavior of $G_{VV}$ component
 \[G_{VV}\approx |V|^{2\beta-2}.\]
 We could have directly figured out this behavior by taking a derivative with respect to $v_2$ of (\ref{mretro}). Therefore, the mass inflation analysis is able to exactly produce the same behavior as reported in \cite{Dias:2019ery}. This behavior captures both the weaker ($\beta<1$) and stronger ($\beta <1/2$) bounds on $\beta$ for which the SCC is respected. In the Christodoulou version of SCC, the perturbing test field should not belong to the space of locally square-integrable functions.\footnote{A function $\phi$ is square-integrable in a domain, if there exists a smooth compactly-supported function $\psi$, such that for $\Phi=\psi\phi,\, (\Phi^2+\partial_{\mu}\Phi\partial_{\mu}\Phi$) is integrable. See \cite{Dias:2018ynt}, for further discussions.}  
 We see, this condition turns out to be the case when
 \be 
 2(\beta-1)<-1 \implies \beta<\frac{1}{2}.
 \ee
 In the following section we use the mass function (\ref{mretro}), and obtain a regular metric in the vicinity of $\mathcal{CH}^+_R$.

\section{A Regular Mass Inflation Solution}\label{regular_soln_section}

We now introduce a set of double null coordinates \cite{Ori:1991zz,Bhattacharjee:2016zof}, $\{U, V, \widetilde{ \phi}\}$, that is regular at $\mathcal{CH}^+_R$ and describes the interior region of the black hole, and then determine the metric functions in the region of mass inflation. The background metric in these new coordinates will take the form \cite{Dias:2019ery, Carlip:1995qv}

\begin{align} \label{double_null_mi}
ds^2=&\Omega^2(U,V) \mathop{dU}\mathop{dV}+r^2(U,V) \nonumber \\ &\left(\mathop{d\widetilde{\phi}+\dfrac{J K(U,V)}{2r^2(U,V)}(V\mathop{dU}-U\mathop{dV})}\right)^2,
\end{align}
where the function $K(U,V)$ should be well behaved at $\mathcal{CH}^+_{R}$. 

We have set up this coordinate system in the following fashion (see Fig. \ref{ori-model_FIGURE}): inside the black hole, $U<0$ and $V<0$, with $\mathcal{CH}^+_R$  is at $V=0$ (and $U<0$). $\mathcal{CH}^+_L$ is at $U=0$. 

The ring $\mathcal{R}$ of radius $R$ is located at a line along which $U$ is constant (a $U=k$ surface where $k$ is a constant), and on the ring, we set $V=\lambda$ where $\lambda$ is the affine parameter. The ring, as earlier, has an intrinsic coordinate $\chi$. So, $R(\lambda):=r(U=k, V=\lambda)$. Thus, according to (\ref{asym_eq}), inside the mass inflation region (region II), $v_2$ is directly proportional to $V$, viz.,
\begin{equation}\label{v=V}
v_2 \propto V.
\end{equation}
Therefore, the mass function goes as
\be
m_2(V)\propto |V|^{-(1-2\beta)}.
\ee 
As already discussed in the earlier section, this result shows the mass inflation occurs if the black hole is sufficiently far from extremality. In the rest of this section, we will assume $\beta <1/2$, for mass inflation to take place and find a regular solution. 

Before proceeding further, we impose a slow-rotation approximation on (\ref{double_null_mi})  in order to get an analytically tractable solution, that is, we keep terms which are of the order $\mathcal{O}(J)$, and obtain the following simplified metric
\begin{align}
ds^2= & \Omega^2(U,V) \mathop{dU}\mathop{dV}+J V K(U,V) \mathop{dU} \mathop{d\widetilde{\phi}} \nonumber \\& -J U K(U,V)\mathop{dV} \mathop{d\widetilde{\phi}}+ r^2 \mathop{d\widetilde{\phi}^2}+ \mathcal{O}(J^2). \label{double_null_g}
\end{align} 

The slow rotation limit will not spoil the generality of our calculation: Since rapidly spinning BTZ black holes are near-extreme, we can trust mass inflation solution in this limit (see Sec. \ref{bound}). 

Using the ingoing EF metric tensor ${g}_{\mu \nu}$ from (\ref{btz_vaidya}), we can use the transformation relation between the two metrics to determine unknown functions $\Omega(U,V)$ and  $K(U,V)$ in (\ref{double_null_mi}); But we must first determine $r$ and $\theta$ as functions of the new coordinates. 

To determine the radial part $r(U,V)$, we recall (\ref{nullity2}), and write 
\begin{equation}
\dfrac{ \partial r}{ \partial V}= \dfrac{N^2}{2} \approx -\dfrac{m_2(V)}{2},
\end{equation}
where, in writing the last relation, have used the fact that during mass inflation, the mass function $m_2(V)$ will grow drastically and hence dominate over the other terms in $N^2$. Employing the mass inflation solution (\ref{v=V}), we can readily integrate this equation (considering the retrograde mode solution) to get 
\begin{equation}
r \approx \gamma |{V}|^{2\beta} +H(U),
\end{equation}
where $\gamma$ is a constant, and $H(U)$ is an arbitrary function that reflects our freedom in choosing the $U$ coordinate. We set 
\begin{equation}
H(U)=r_{-}- \epsilon \abs{U},
\end{equation}
where $\epsilon$ is a dimensionless constant representing the strength of the outflux \cite{Balbinot:1993rf}, and get
\begin{equation} \label{r(U,V)1}
r \approx r_{-}- \epsilon |U| + \gamma|{V}|^{2\beta}.
\end{equation}
The above equation holds near $\mathcal{CH}^+_R$ and becomes finite at $V=0$. We choose $\epsilon |U|$ with a minus sign as it reflects that $\mathcal{CH}^+_R$ will contract with time because of the focusing effect of the generators of $\mathcal{CH}^+_R$.

Now recalling the geodesic equation for the $\theta$ coordinate (\ref{theta_dot_eq}), we can write
\begin{equation}
\dfrac{\partial \theta}{ \partial V}=\dfrac{J}{2r^2}.
\end{equation}
This equation may be integrated in principle to write the coordinate $\theta$ in terms of $U$,$V$ and $\widetilde{\phi}$ up to an arbitrary function $f(U)$, but we note from the consistency of the transformation relation between the metrics, this $f(U)$ can be set to zero.
Now, again using the transformation relations, we find
\begin{equation} \label{omega2}
\dfrac{\Omega^2(U,V)}{2} \approx \dfrac{\partial r}{\partial U} \approx \dfrac{\Omega^2_0}{2},
\end{equation}
where $\Omega_0$ is a constant. The function $K(U,V)$ can be determined likewise but for our purpose, it is just important to note that $K(U,V)$ and its derivatives are well behaved at $\mathcal{CH^+_R}$. This completes the determination the of a regular mass inflation solution at $\mathcal{CH}^+_R$. Although this solution is regular at the $\mathcal{CH}^+_R$ but an unbounded growth of the quasilocal Hawking mass may still be obtained with this regular metric as described in \cite{Poisson:1989zz,Poisson:1990eh}. 

\subsection{Tidal distortions and nature of the singularity}\label{tidal}

We now consider an observer inside the black hole, and we shall try to understand what they are likely to experience as they approach $\mathcal{CH}^+_R$. We note that the following analysis will be more accurate if our observer approaches the early portion of $\mathcal{CH}^+_R$. We shall model this extended observer as a collection of pointlike particles moving along timelike geodesics. As the observer approaches the mass inflation singularity, they will experience strong tidal forces and suffer distortions. To study the local tidal effects near $\mathcal{CH}^+_R$, we set up a following orthonormal set of triads that satisfy
\begin{equation}\label{traid_condition}
\nabla_{\nu}\hat{e}^{\mu}_{(\alpha)}u^{\nu}=0,~ 
\hat{e}^{\mu}_{(0)}=u^{\mu},~\mathrm{and}~
\hat{e}^{\mu}_{(\alpha)} g_{\mu \nu} \hat{e}^{\nu}_{(\beta)} = \eta_{(\alpha)(\beta)}
\end{equation}
near $\mathcal{CH}^+_R$, where $u^{\mu} = d x^{\mu}/d \tau$ is the velocity of the observer in the background (\ref{double_null_g}) with which they approach $\mathcal{CH}^+_R$ ($\tau$ being the proper time). The orthonormal set of triads is given by

\begin{equation}
\begin{matrix}
{\hat{e}}^{\mu}_{(0)}=& \dfrac{1}{{\Omega_0}}\bigg(& \dfrac{-1}{u^V}, & u^V, & 0&\bigg),\\[2ex]
{\hat{e}}^{\mu}_{(1)}=&  \dfrac{1}{{\Omega_0}}\bigg(& \dfrac{1}{u^V}, & u^V, & 0&\bigg),\\[2ex] 
{\hat{e}}^{\mu}_{(2)}=&\dfrac{1}{{\Omega_0}}\bigg(&\frac{JUK(r)}{\Omega_0 r}, & -\frac{JVK(r)}{\Omega_0 r}, & \dfrac{\Omega_0}{r} &\bigg).\\
\end{matrix}
\end{equation}

Now we define a small separation vector $\xi^{\mu}$, connecting any particle that make up our observer to its center of mass, as $\xi^{\mu}=\bar{x}^{\mu}(\tau)-x^{\mu}(\tau)$. The geodesic deviation equation tells us that it evolves according to the equation
\begin{equation}\label{geodesicdev_eq}
\dfrac{D^2 \xi^{\mu}}{D \tau^2}=-R^{\mu}_{~\alpha \beta \gamma}u^{\alpha}\xi^{\beta}u^{\gamma}.
\end{equation}
Since the observer is a spacelike entity, we shall project $\xi^{\mu}$ along the spacelike dual vectors and define
\begin{equation}
\bar{\zeta}^{i}=\hat{e}^{(i)}_{\mu} \xi^{\mu} 
\quad \quad\mathrm{where} \quad 
i=1,2
\end{equation} 
Now, noting that we are in a locally inertial frame, (\ref{geodesicdev_eq}) gives us
\begin{equation}\label{tidal_d_eq}
\dfrac{d^2 \bar{\zeta}^{i}}{d \tau^2}+K^i_j \bar{\zeta}^j=0,
\end{equation}
where
\begin{equation}
K_{ij}=R_{\alpha \beta \gamma \delta}e^{\alpha}_i u^{\beta}e^{\gamma}_j u^{\delta}.
\end{equation}
We can readily calculate the components of $K_{ij}$ near $\mathcal{CH}^+_R$, in the slow rotation limit, and write
\begin{flalign}
	K_{11}& \approx \mathcal{O}(J^2), \\
	K_{22}& \approx  -\dfrac{{(u^V)}^2 }{\Omega_0^2 r_{-}}\partial^2_V r \Big{|}_{r \to r_{-}} + \mathcal{O}(J^2).
\end{flalign} 
This $K_{22}$ component shall reveal the strength of the tidal forces. Now as the geodesic equation for $u^V$ is simply
\begin{equation}
\dfrac{du^V}{d \tau}=0
\quad \text{(since $u^{\phi}=0$ by construction)},  \qquad
\end{equation}
we can say that $u^V$ is a constant and (recalling the definition of $u^V$) $V \propto \tau$ near $V=0$. Using this we see that the leading divergence in $K_{22}$ near  $\mathcal{CH}^+_R$ is

\begin{equation}
	K_{22} \propto \frac{\abs{V}^{2(\beta -1)}} {r_-}.
\end{equation}
Plugging this into (\ref{tidal_d_eq}), and integrating it twice, we can, therefore, see that the tidal distortions remain finite in the limit $V \to 0$, even though the tidal forces diverge near  $\mathcal{CH}^+_R$. Since the distortion is finite, we can say that the mass inflation singularity is a weak null singularity (remembering that the metric components are finite at $\mathcal{CH}^+_R$) \cite{Ori:1991zz} . So, our observer may safely cross the $\mathcal{CH}^+_R$ and emerge in a new universe in the classical picture. 

Therefore, we conclude that a rotating BTZ black hole respects the strong cosmic censorship conjecture. In this mass inflation version, the metric is regular at  $\mathcal{CH}^+_R$, but its derivatives cannot be smoothly extended across $\mathcal{CH}^+_R$. 

\subsubsection{Bounds on the spin parameter}\label{bound}
It will be a good exercise to find an exact number in terms of the parameters of a BTZ black hole such that one has a rough idea about when SCC is violated, or find a bound on the parameters beyond which we can say the black hole is sufficiently far from extremality. Recently, by solving an Einstein-Klein Gordon system in $AdS_3$ space using numerical methods, \cite{Pandya:2020ejc} has shown that the spin parameter $a$ of BTZ black hole should be less than $0.6$ for which SCC or mass inflation is respected. Their result is consistent with that of \cite{Dias:2019ery}. Recall, for $\beta<1$, we get the weaker version of SCC from the behavior of stress-energy tensor (\ref{stfalloff}). Setting $\beta\leq 1$, for massless field we get $r_+\geq 3r_-$. For $r_+\geq 3r_-$, SCC is respected. Now the spin of the black hole can be expressed as 
\be
a=\dfrac{J}{M L}=\dfrac{2r_+r_-}{r_+^2~+~r_-^2}=\dfrac{2r_+r_-}{(r_+-r_-)^2+2r_+r_-}.\label{spin}
\ee
Putting the largest value of $r_+=3r_-$, such that our model respects SCC, gives
\be
a=\dfrac{3}{5}=0.6,
\ee
and if we try to find the bound for the stronger case, $\beta<1/2$, we get a stricter bound \be a<0.38.\ee 

Although this stronger bound has not been explicitly shown in \cite{Pandya:2020ejc} but our analysis shows that there is a region in the parameter space when Einstein equation cannot be extended even as weak solutions beyond $\mathcal{CH}^+_R$. We can also justify from these bounds that our slow rotation approximation in obtaining a solution for mass inflation region is not a crude assumption. Looking at the last expression of $a$, one can easily see, that near-extremal black holes are rapidly spinning ones. So small spin (or slow rotation) is consistent with the singular nature of $\mathcal{CH}^+_R$.

\subsection{Quantum effects at $\mathcal{CH}^+_R$}\label{backreaction}

As the classical backreaction in a BTZ spacetime has produced a singularity at $\mathcal{CH}^+_R$, it is expected that quantum backreaction will also produce a similar effect. Of course, if one gets close to the extremality, the classical analysis shows a breakdown of SCC. It is interesting to learn from \cite{Dias:2019ery} that quantum corrections cannot remove the classical breakdown. This breakdown has also been explained in \cite{Hollands:2019whz}, and is supported from different perspectives in \cite{Balasubramanian:2019qwk, Papadodimas:2019msp}. However, as argued in \cite{Emparan:2020rnp}, this may not be true if one goes beyond linear perturbations. It seems, at least for linear order perturbations, there is no strong backreaction effects at $\mathcal{CH}^+_R$.  In their study, Dias \textit{et al.} \cite{Dias:2019ery} have shown, for $\beta <1$ the expected value of stress-energy tensor $\braket{0|T_{\mu\nu}|0}$ diverges near $\mathcal{CH}^+_R$, but indicated that the divergence is integrable. In \cite{Hollands:2019whz},  it has been shown that the component $T_{VV}$ of renormalized stress-energy tensor in a state $\psi$, diverges in the following universal fashion for a wide class of theories in some Kruskal like coordinate system near $\mathcal{CH}^+_R$
\be
\braket{T_{VV}}_{\psi}\approx \dfrac{C}{V^2}~+~t_{VV}.
\ee
Here $C$ is a constant that depends only on the black hole parameters, and not on $\psi$. The divergence of $t_{VV}$ depends on $\psi$, but it diverges no more strongly than the stress-energy tensor of a classical solution. The analysis of \cite{Dias:2019ery} shows $C$ is zero for a BTZ black hole in the near-extreme zone for Hartle-Hawking state. In \cite{Hollands:2019whz}, it was also explicitly shown this will be the case for a BTZ black hole. We believe similar situation shall arise as long as we are confined within the linear order in perturbations. Now, let us try to see the behavior of $t_{VV}$ for a BTZ black hole. If we calculate the $VV$ component of the Einstein tensor near $\mathcal{CH}^+_R$  taking a quantum backreacted metric as in (\ref{double_null_mi}), we get the following semiclassical Einstein equation
 
\begin{equation}
G_{VV}\simeq\frac{(2 \partial_V\Omega \partial_V r  + \Omega \partial^2_V r )}{r \Omega}=-8 \pi t_{VV},\label{bkrceq}
\end{equation}
where we have only considered that $J$ is small. This implies we are far from extremality.  In this zone, if the divergent behavior of $t_{VV}$ has to be as bad as that of classical singularity \cite{Hollands:2019whz}, then  plugging in the expression of $r$ from (\ref{r(U,V)1}) we get

\begin{equation}
    t_{VV}\approx \dfrac{|V|^{2\beta-2}}{r_-}.
\end{equation}

This is consistent with the behavior of $t_{VV}$ suggested in \cite{Hollands:2019whz}. In this heuristic analysis, we see if we are far from extremality, or $\beta<1$, there will be a mild divergence in the quantum stress-energy tensor as indicated in \cite{Dias:2019ery}. 

\section{Conclusion }\label{discussion}
We have made an analytical study of mass inflation in a rotating BTZ black hole. Using the Ori model to implement mass inflation, we have shown how the right Cauchy horizon ($\mathcal{CH}^+_R$) becomes unstable for a nonextreme BTZ black hole. The important aspect of our analysis is to show, in the mass inflation model, instead of the prograde QNMs, it is the faster decaying retrograde QNMs that provide the correct singular structure of $\mathcal{CH}^+_R$ as depicted in \cite{Dias:2019ery}. It is remarkable to obtain the correct behavior at $\mathcal{CH}^+_R$ with a simple toy model, that too with a faster mode. We have also obtained a regular solution in the mass inflation region. To test the strength and nature of the singularity, we constructed a freely falling frame and calculated the tidal forces employing a slow rotation approximation. The tidal forces become divergent as we approach $\mathcal{CH}^+_R$ but their twice integrated values (tidal distortions) remain finite, a signature of a weak null singularity. This feature of a weak singularity is restricted to a body that is sufficiently small. For a large object like a planet, the geodesic deviation equation would contain higher order derivatives of the metric, and the strength of the singularity may not be weak. This may be an interesting thing to study further. We have also reproduced some of the bounds on the black hole spin parameter $a$ from different restrictions on $\beta$ so that SCC is respected. The weaker bound, $a<0.6$, seems to suggest a rapidly spinning black hole is close to its extremality, and it is more prone to violate SCC.  We also indicate a stronger bound, $a<0.38$, that will rule out the existence of any weak solution of the Einstein equations and protect SCC. Finally, we have given a heuristic argument to show that the behavior of quantum stress-energy tensor conjectured in \cite{Dias:2019ery,Hollands:2019whz} appears to be true for the BTZ black hole. It will be interesting to see numerically, whether the mild divergence of $t_{VV}$ remains true for a BTZ black hole sufficiently far from extremality .

The mass inflation model should be also able to reproduce the recent studies related to SCC violation for near-extreme de Sitter black holes. For BTZ and other higher dimensional $AdS$ black holes, it will be interesting to see the perturbative effects of fields other than the scalar field.  These studies may lead to new insights in matters relates to strong cosmic censorship. In a $4D$ $AdS$ black hole, due to stable trapping phenomena, we get a much slower logarithmic decay of the perturbations outside and along the event horizon. This will induce a more rapid divergence at the Cauchy horizon. It will be interesting to see whether the universal form of stress-energy tensor suggested in \cite{Hollands:2019whz} works there as well or not.

\section{Acknowledgments} \label{acknowledgment}
 S.B. thanks Amitabh Virmani for many suggestions and useful inputs in this project. The authors also thank Amitabh Virmani for reading through an earlier version of this manuscript. Research of S.B. and S.S. is supported by DST-SERB, Government of India under the scheme Early Career Research Award (File no.: ECR/2017/002124) through the project titled \emph{``Near Horizon Structure of Black Holes"}. The authors would like to also acknowledge support from IIIT, Allahabad through the Seed Grant for the project \emph{``Probing the Interior of $AdS$ Black Holes"}. The computer algebra system \emph{Wolfram Mathematica} has been used to check certain calculations.

%\bibliographystyle{abbrv}
%\bibliographystyle{plain}
%\bibliographystyle{JHEP}
%\bibliographystyle{apsrev}
%\bibliography{BTZMF}

%\iffalse  %uncomment to ignore the following bibitems 


%\fi uncomment if \ifflase is uncommented

\end{document}